\documentclass[aip,reprint]{revtex4-1}

\draft 

\usepackage{graphicx}
\usepackage{dcolumn}
\usepackage{bm}

\usepackage[utf8]{inputenc}
\usepackage[T1]{fontenc}
\usepackage{mathptmx}

\begin{document}


\title{High-numerical-aperture and long-working-distance objectives for single-atom experiments} 



\author{Shaokang Li}
\author{Gang Li}
\email[]{gangli@sxu.edu.cn}
\author{Wei Wu}
\author{Qing Fan}
\author{Yali Tian}
\author{Pengfei Yang}
\author{Pengfei Zhang}
\author{Tiancai Zhang}
\email[]{tczhang@sxu.edu.cn}
\affiliation{State Key Laboratory of Quantum Optics and Quantum Optics Devices, and Institute of Opto-Electronics, Shanxi University, Taiyuan 030006, China}
\affiliation{Collaborative Innovation Center of Extreme Optics, Shanxi University, Taiyuan 030006, China}


\date{\today}

\begin{abstract}
We present two long-working-distance objective lenses with numerical apertures (NA) of 0.29 and 0.4 for single-atom experiments. The objective lenses are assembled entirely by the commercial on-catalog $\Phi$1'' singlets. Both the objectives are capable to correct the spherical aberrations due to the standard flat vacuum glass windows with various thickness. The working distances of NA$=0.29$ and NA$=0.4$ objectives are 34.6 mm and 18.2 mm, respectively, at the design wavelength of 852 nm with 5-mm thick silica window. In addition, the objectives can also be optimized to work at diffraction limit at single wavelength in the entire visible and near infrared regions by slightly tuning the distance between the first two lenses. The diffraction limited fields of view for NA$=0.29$ and NA$=0.4$ objectives are 0.62 mm and 0.61 mm, and the spatial resolutions are 1.8 $\mu$m and 1.3 $\mu$m at the design wavelength. The performances are simulated by the commercial ray-tracing software and confirmed by imaging the resolution chart and a 1.18 $\mu$m pinhole. The two objectives can be used for trapping and manipulating single atoms of various species.
\end{abstract}


\maketitle 


\section{Introduction}
As one of the most important tasks in quantum research field, quantum computing \cite{Ladd2010} has been explored for decades based on various of physical systems. Optically trapped neutral single-atoms system, which has the advantages of the long coherence time and good scalability, has been regarded as one of the most promising systems to achieve the quantum computation \cite{Saffman2018, Weiss2017, Saffman2010}. In the very recent time, the demonstration of high fidelity quantum gates via Rydberg interaction accelerated the researches on the neutral atom-based quantum computation \cite{Levine2019, Graham2019}. The foundation of the natural atom-based quantum computing is the resolvable single atom array \cite{Kumar2018, Labuhn2016, Barredo2018, Endres2016, Xia2015, Piotrowicz2013, Nelson2007}. The objective lens systems, which possess diffraction limited condition within a large field of view (FOV), high numerical aperture (NA), and long working distance (WD), are used in many experiments \cite{ALT2002, Weber2006, Nelson2007, Piro2011, Zimmermann2011, Piotrowicz2013} to form the micro-sized optical dipole trap array and load single atoms from magneto-optical trap (MOT). The same objective lens is usually adopted to observe the trapped single atoms. In order to be compatible with specific atomic species and vacuum chambers, the objective lenses are always custom designed and manufactured. This made them costly and time-consuming to be obtained.

Many substituting designs by mostly using the on-catalog commercial spherical singlet appeared in recent years. The first one is the ``Alt objective'' \cite{ALT2002}, which uses three pieces of commercial singlets and one piece of custom-designed singlet to realize a diffraction limited objective with NA$=0.29$. All the singlets have size of $\Phi$1'' in this design so that the objective is compatible to the commonly used $\Phi$1'' optics in experiment. Later, several designs with all commercial $\Phi$2'' singlets were reported and the diffraction limited NA is between 0.175 and 0.44 upon different designs  \cite{Li2018, Pritchard2016, Bennie2013}. All these designs provide more options for the single-atom experiments. 

In this paper, we provide two additional designs of objective lenses with all $\Phi$1'' on-catalog commercial singlets from Thorlabs Inc. Both objectives can correct the spherical aberrations due to the standard flat vacuum glass windows with various thickness. The NAs for the two objectives are NA$=0.29$ and NA$=0.4$ with WDs of 34.6 mm and 18.2 mm, respectively, at design wavelength of 852 nm with a standard 5mm-thick silica vacuum window. These two objectives can be optimized to work at diffraction limited condition for various thickness of vacuum window under single wavelength in the entire visible and near infrared regions by slightly tuning one lens spacing. The diffraction limited FOV for NA$=0.29$ and NA$=0.4$ objectives are 622 $\mu$m and 610 $\mu$m. The resolutions for the two objectives are 1.8 $\mu$m and 1.3 $\mu$m at the design wavelength. The performances are simulated by the commercial ray-tracing software and confirmed by imaging the resolution chart and a 1.18 $\mu$m pinhole. The two objectives can be used for trapping and manipulating single atoms of various species.

\section{NA$=0.29$ objective lens assembly}

\begin{figure}
\includegraphics[width=8.5cm]{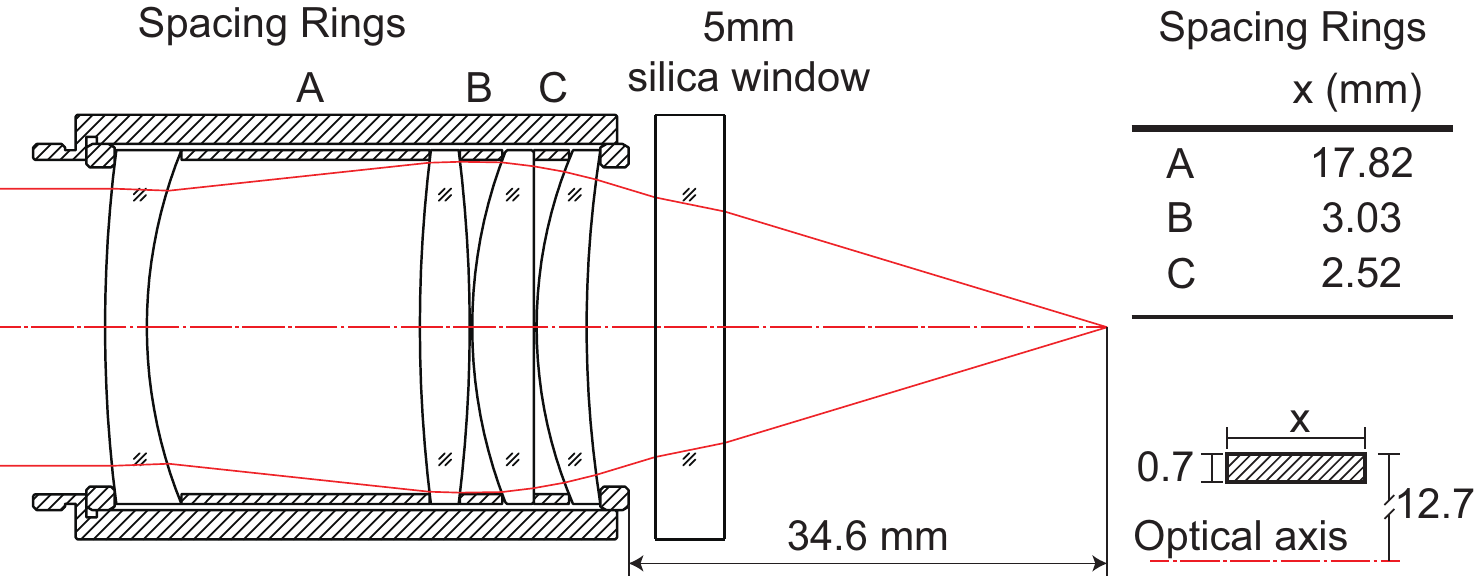}
\caption{\label{fig1} Structure of the NA$=0.29$ objective lens assembly. The four on-catalog commercial lenses from left to right are LF1822, LB1676, LA4725, and LE1234 (Thorlabs). A, B, and C are three custom-made spacing rings.}
\end{figure}

The structure of NA$=0.29$ lens assembly is shown in Fig. \ref{fig1}. Four commercial catalog singlets (LF1822, LB1676, LA4725, and LE1234 from Thorlabs) are mounted in the standard commercial $\Phi$1'' lens tube (SM1L10, Thorlabs) with three custom-made spacing rings. The drawing and size of the rings are also depicted in Fig. \ref{fig1}. Two retaining rings (SM1RR, Thorlabs) are used to fasten the whole lens assembly. The complete prescription of the objective designed for 5-mm-thick silica vacuum window at 852 nm is given in Table \ref{tab1}. The whole structure is similar to the Alt lens \cite{ALT2002} excepting that the first lens is a negative Meniscus lens. The Alt lens uses three commercial on-catalog lenses and one custom-made positive Meniscus lens. Our design ensures the objective assembled from the four commercial catalog singlets working at the diffraction limited while keeping the maximum NA. The effective focal length (EFL) is bout 34.4 mm, and the WD of the objective is about 34.6 mm. 

\begin{table}[ht]
\caption{\label{tab1} NA$=0.29$ objective prescription. The design wavelength is 852 nm. VW means vacuum window. }
\centering
\begin{tabular}{ccccc}
\hline \hline
Surface & Curvature(mm) & Thickness (mm) & Material & Lens\\
\hline
1 & 100 & 3 & NBK7 & LF1822\\
2 & 33.7 & 19.6 (d) & air & \\
3 & 102.4 & 3.6 & NBK7 & LB1676\\
4 & $-102.4$ & 0.2 & air & \\
5 & 34.5 & 4.4 & Silica & LA4725\\
6 & $\infty$ & 0.2 & air & \\
7 & 32.1 & 3.6 & NBK7 & LE1234\\
8 & 82.2 & 5 & air & \\
9 & $\infty$ & 5 (D) & Silica & VW\\
10 & $\infty$ & 27.65 & vacuum & \\
\hline
\end{tabular}
\end{table}

\begin{figure*}[hbt!]
\includegraphics[width=17cm]{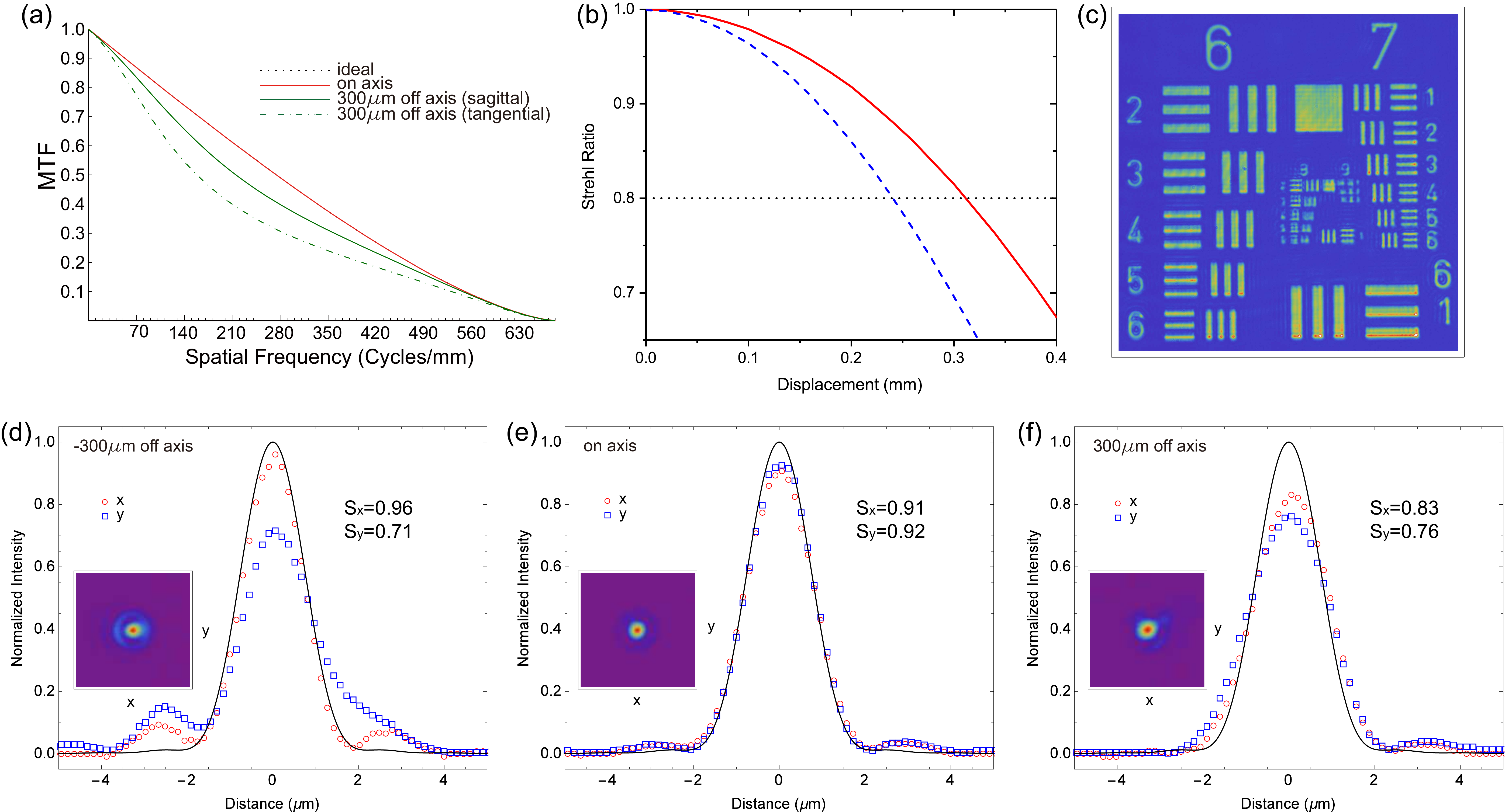}
\caption{\label{fig2} The performance of the NA=0.29 objective lens. (a) The simulated modulation transfer function (MTF) for focal spots on axis as well as 300 $\mu$m off axis and the ideal diffraction limited MTF. (b) Calculated Strehl ratio. The red-solid and blue-dashed lines are the results for the original design and the design with the plano-convex lens LA4725 replaced by LA1608. The diffraction limited field of view (FOV) is defined as the displacement where Strehl ratio $\geq 0.8$ (the dashed line). (c) Image of USAF 1951 resolution target. (d)-(f) The measured point spread function (PSF) by imaging a 1.18-$\mu$m pinhole at the positions of -300, 0, and 300 $\mu$m off the optical axis. The red circles and the blue squares are the normalized intensity distribution along x-axis and y-axis which across the intensity maxima, whereas the black curve is the convolution of a 1.18 $\mu$m top-hat function with the point spread function. The Strehl ratios along the two axes are given by S$_x$ and S$_y$.  Insets of (d)-(f) are the images of the 1.18-$\mu$m pinhole at the corresponding positions. }
\end{figure*}

The performance of the objective is optimized and simulated by the commercial ray-tracing software (Zemax). The comparison of the simulated modulation transfer function (MTF) for focal spots on axis and 300 $\mu$m off axis to the diffraction limited MTF is shown in Fig. \ref{fig2}(a). The calculated Strehl ratio versus off-axis displacement is displayed in Fig. \ref{fig2}(b) as the solid red line. We can see that the MTF of our objective is almost overlap with the ideal one and the Strehl ratio is 1 when the focal spot is on axis, which means that our objective lens has corrected the aspheric errors perfectly. The off-axis performance is a little bit worse. The diffraction limited FOV is usually defined as the displacement from the axis where Strehl ratio $\geq 0.8$ \cite{Gross2015}. The theoretical FOV for our objective is 0.622 mm. 

To experimentally verify the performance of the objective, a USAF 1951 resolution target (55-622, Edmund Optics) and a 1.18 $\mu$m pinhole (P1H, Thorlabs) are used to evaluate the objective by imaging them on a CCD camera (UC500M, CatchBest) with a $f=500$ mm achromatic lens (AC254-500-B, Thorlabs). A collimated 852-nm laser beam is used to illuminate the objects and a 5 mm fused silica flat was adopted to imitate the vacuum window. The 500-mm achromatic lens is placed right after the objective and the CCD camera is placed at the focal point of the achromatic lens.

The image of the resolution target is shown in Fig. \ref{fig2}(c) and the magnification of the imaging system can be determined as $-14.53$ from the image by considering the pixel size (2.2 $\mu$m $\times$ 2.2 $\mu$m) of the camera. We can see that the line pairs in element 2 of group 8 can be clearly resolved, this means a resolution about 1.7 $\mu$m can be achieved. The result is consistent with the theoretical resolution $0.61 \lambda / \text{NA}= 1.8$ $\mu$m. Next, by imaging the 1.18 $\mu$m pinhole the point spread function (PSF) is measured. Fig. \ref{fig2}(d)-(f) show the results with the pinhole placed -300, 0, and 300 $\mu$m off the optical axis. The insets of each figures are the corresponding images of the pinhole. The measured intensity distribution along x-axis and y-axis are shown as red open circles and blue open squares, respectively. The black solid curves are the theoretical intensity distribution with an ideal lens system, and the curve is given by the convolution between a 1.18 $\mu$m top-hat function and the ideal PSF. The theoretical curve are normalized to the maximum intensity, while all the experimental data are normalized as the overall power same to the theoretical one. Therefore, the maximum intensity on the experimental data represents the measured Strehl ratio \cite{Gross2015}, and they are shown as $S_x$ and $S_y$ for the two directions in each figure. We can see that the average Strehl ratios for the three positions of the pinhole are no less than 0.8. The results confirm that the diffraction limited FOV of our objective lens is greater than 0.6 mm.

\begin{figure}[hbt!]
\includegraphics[width=8.5cm]{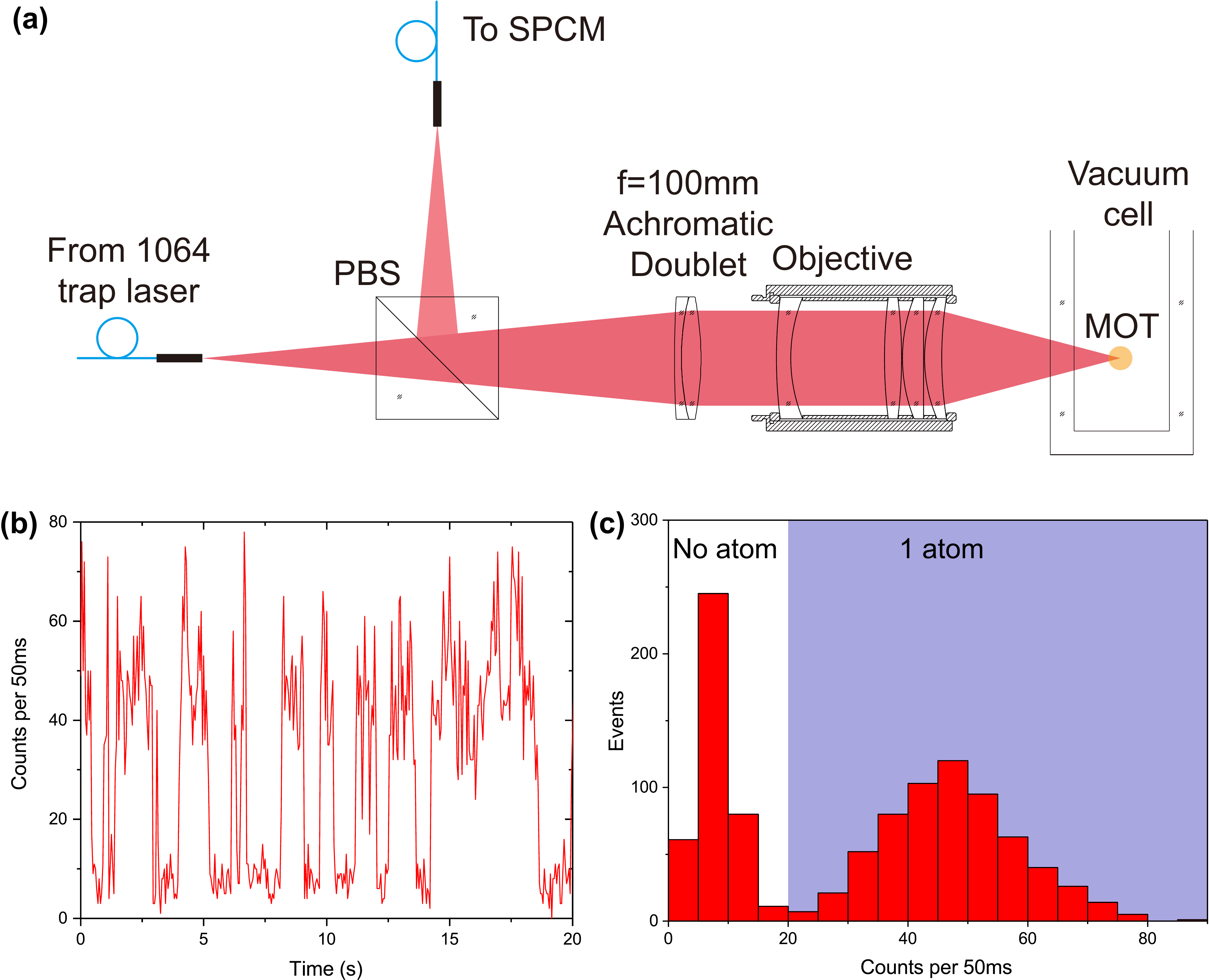}
\caption{\label{fig3} Single atom loading and observation by using the NA=0.29 objective lens. (a) The sketch of the optical system to produce the dipole trap and observation of the loaded single atoms. (b) The recorded single atom signal. (c) The histogram of the single atom signal.}
\end{figure}

This objective lens has been used in our lab to generate a micro-sized dipole trap to capture single atoms from a MOT. Fig. \ref{fig3}(a) is the sketch of the trap formation and observation systems. A 1064-nm laser light from a single mode fiber is collimated by an achromatic doublet (AC254-100-B, Thorlabs) and then focused by the objective lens to a spot size about 2.1 $\mu$m. When the position of the trap overlaps MOT, the single atoms can survive in the trap due to the light-assistant two-body collisions \cite{Schlosser2001, Schlosser2002}. The scattered photons from the single trapped atom is then collected by the same objective lens and focused by the same 100-mm achromatic doublet. After the reflection of a polarization beam splitter, the s-polarized photon is coupled into another optical fiber and sent to the single photon counting modules (SPCMs). Fig. \ref{fig3}(b) is the recorded signals, and (c) is the histogram. These two figures show clearly that only one atom can be loaded. 

Actually, the third plano-concave silica lens (LA4725) can be replaced by the counterpart made from N-BK7 (LA1608). In this case, the distance between the first two lenses should be optimized to d=23.5 mm to get the diffraction limited condition. The corresponding performances are similar except that a little smaller FOV. The calculated Strehl ratio versus off-axis displacement is also displayed in Fig. \ref{fig2}(b) as blue dashed line. The theoretical FOV is then 0.48 mm.

\begin{table}[ht]
\caption{\label{tab2} (a) NA$=0.29$ objective optimized for various wavelength with the window thickness D=5 mm. (b) NA$=0.29$ objective optimized for various window thickness D at the wavelength of 852 nm. NA is kept as 0.29 for all the optimizations. $\lambda$, d, and D are the working wavelength, distance between the first two lenses (thickness of surface 2 in Table \ref{tab1}), and window thickness, respectively.}
\centering
\begin{tabular}{l}
(a) \\
\begin{tabular}{cccc}
\hline \hline
$\lambda$ (nm) & d (mm) & EFL (mm) & FOV (mm)\\
\hline
405 & 18.36 & 33.2 & 0.22\\
436 & 18.59 & 33.4 & 0.25\\
486 & 18.86 & 33.6 & 0.29\\
532 & 19.03 & 33.8 & 0.34\\
589 & 19.17 & 34.0 & 0.38\\
671 & 19.34 & 34.1 & 0.46\\
780 & 19.50 & 34.3 & 0.55\\
852 & 19.60 & 34.3 & 0.62\\
1060 & 19.82 & 34.5 & 0.80\\
1550 & 20.17 & 34.9 & 1.16\\
\hline
\end{tabular}
\\
\\
(b) \\
\begin{tabular}{cccc}
\hline \hline
D (mm) & d (mm) & EFL (mm) & FOV (mm)\\
\hline
9 & 21.45 & 33.7 & 0.82\\
8 & 20.95 & 33.8 & 0.77\\
7 & 20.50 & 34.0 & 0.72\\
6 & 20.00 & 34.2 & 0.67\\
5 & 19.60 & 34.3 & 0.62\\
4 & 19.16 & 34.5 & 0.58\\
3 & 18.74 & 34.7 & 0.54\\
2 & 18.36 & 34.8 & 0.51\\
1 & 17.96 & 35.0 & 0.48\\
0 & 17.59 & 35.1 & 0.45\\
\hline
\end{tabular}
\end{tabular}
\end{table}

In addition to the diffraction limited performance at 852nm and window thickness of 5 mm, the objective can also be optimized to diffraction limit at various working wavelengths and for various thicknesses of flat optical window by only adjusting the distance between the first two lenses d (thickness of surface 2 in Table \ref{tab1}). The NA is kept as 0.29 for all the optimization. Table \ref{tab2} (a) and (b) give the results of the optimization.

\section{NA$=0.4$ objective lens assembly}

\begin{figure}[hbt!]
\includegraphics[width=8.5cm]{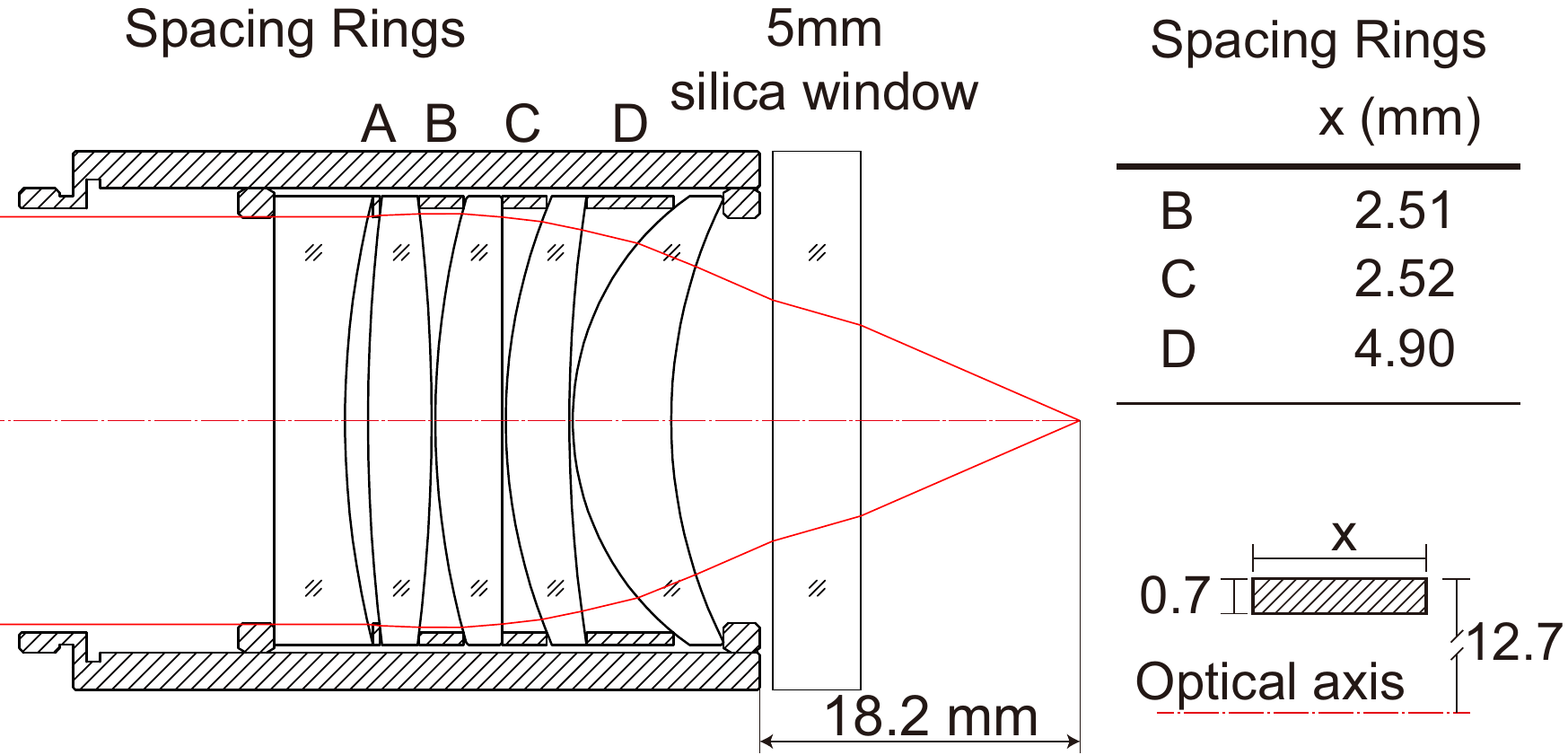}
\caption{\label{fig4} Structure of the NA$=0.4$ objective lens assembly. The five lenses from the left to right are LB1120, LB1676, LA4380, LE1234, and LE5802 from Thorlabs. A is a standard 0.4-mm plastic spacer SM1S01 from Thorlabs. B, C, and D are three custom-made spacing rings.}
\end{figure}

The structure of NA$=0.4$ lens assembly is shown in Fig. \ref{fig4}. Five commercial catalog singlets (LB1120, LB1676, LA4380, LE1234, and LE5802 from Thorlabs) are housed in the standard commercial $\Phi$1'' lens tube (SM1L10, Thorlabs) with 3 custom-made spacing rings (B, C, and D) and a 0.4-mm plastic spacer (A: SM1S01, Thorlabs). The drawing and size of the rings are also shown in Fig. \ref{fig4}. Two retaining rings (SM1RR, Thorlabs) are used to fasten the assembly. The complete prescription of the objective designed for 5mm-thick silica vacuum flat window at 852 nm is given in Table \ref{tab3}. Different from the NA$=0.29$ objective lens, the first lens is a negative plano-concave lens. In order to increase the NA two positive Meniscus lenses are used with focal lengths of 100 mm (LE1234) and 75 mm (LE5802). These promise the objective assembled from the five commercial catalog singlets working at the diffraction limit while keeping the maximum NA=0.4. The EFL is about 28.9 mm, and the WD is about 18.2 mm. 

\begin{table}[ht]
\label{tab3} 
\caption{NA$=0.4$ objective prescription. The design wavelength is 852 nm. VW means vacuum window. }
\centering
\begin{tabular}{ccccc}
\hline \hline
Surface & Curvature(mm) & Thickness (mm) & Material & Lens\\
\hline
1 & $\infty$ & 4 & NBK7 & LB1120\\
2 & 51.5 & 1.3 (d) & air &\\
3 & 102.4 & 3.6 & NBK7 & LB1676\\
4 & $-102.4$ & 0.2 & air &\\
5 & 46 & 3.8 & silica & LA4380\\
6 & $\infty$ & 0.2 & air &\\
7 & 32.1 & 3.6 & NBK7 & LE1234\\
8 & 82.2 & 0.2 & air &\\
9 & 15.5 & 5.6 & CaF$_2$ & LE5802\\
10 & 28.0 & 5.0 (D) & air &\\
11 & $\infty$ & 5 & silica & VW\\
12 & $\infty$ & 13.17 & vacuum &\\
\hline
\end{tabular}
\end{table}

\begin{figure*}[hbt!]
\includegraphics[width=17cm]{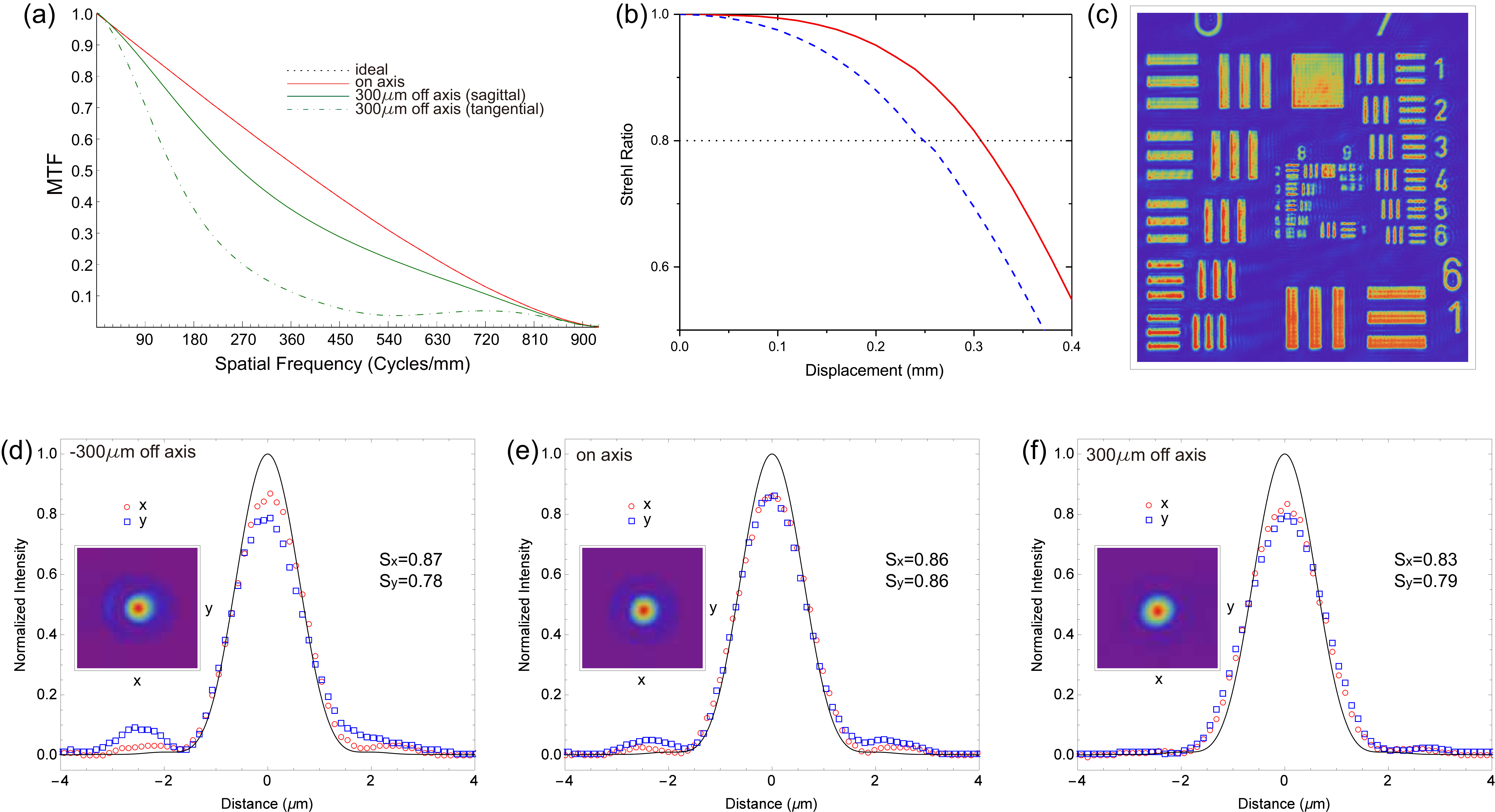}
\caption{\label{fig5} The performance of the NA=0.4 objective lens. (a) The simulated modulation transfer function (MTF) for focal spots on axis as well as 300 $\mu$m off axis and the ideal diffraction limited MTF. (b) Calculated Strehl ratio. The red-solid and blue-dashed lines are the results for the original design and the design with the plano-convex lens LA4380 replaced by LA1509. The diffraction limited field of view (FOV) is defined as the displacement where Strehl ratio $\geq 0.8$ (the dashed line). (c) Image of USAF 1951 resolution target. (d)-(f) The measured point spread function (PSF) by imaging a 1.18-$\mu$m pinhole at the positions of -300, 0, and 300 $\mu$m off the optical axis. The red circles and the blue squares are the normalized intensity distribution along x-axis and y-axis which across the intensity maxima, whereas the black curve is the convolution of a 1.18 $\mu$m top-hat function with the point spread function. The Strehl ratios along the two axes are given by S$_x$ and S$_y$.  Insets of (d)-(f) are the images of the 1.18-$\mu$m pinhole at the corresponding positions.}
\end{figure*}

The performance of this objective is also optimized and simulated by the commercial ray-tracing software (Zemax). The comparison of simulated MTF for focal spots on axis and 300 $\mu$m off axis to the ideal MTF is shown in Fig. \ref{fig5}(a). The calculated Strehl ratio versus off-axis displacement is displayed in Fig. \ref{fig5}(b) as red-solid curve. The overlap of the simulated MTF on axis with the ideal one and the corresponding Strehl ratio with 1 indicate that this objective lens has also corrected the aspheric aberration perfectly. The analyze of Strehl ratio gives the FOV about 0.61 mm. 

\begin{figure}[hbt!]
\includegraphics[width=8.5cm]{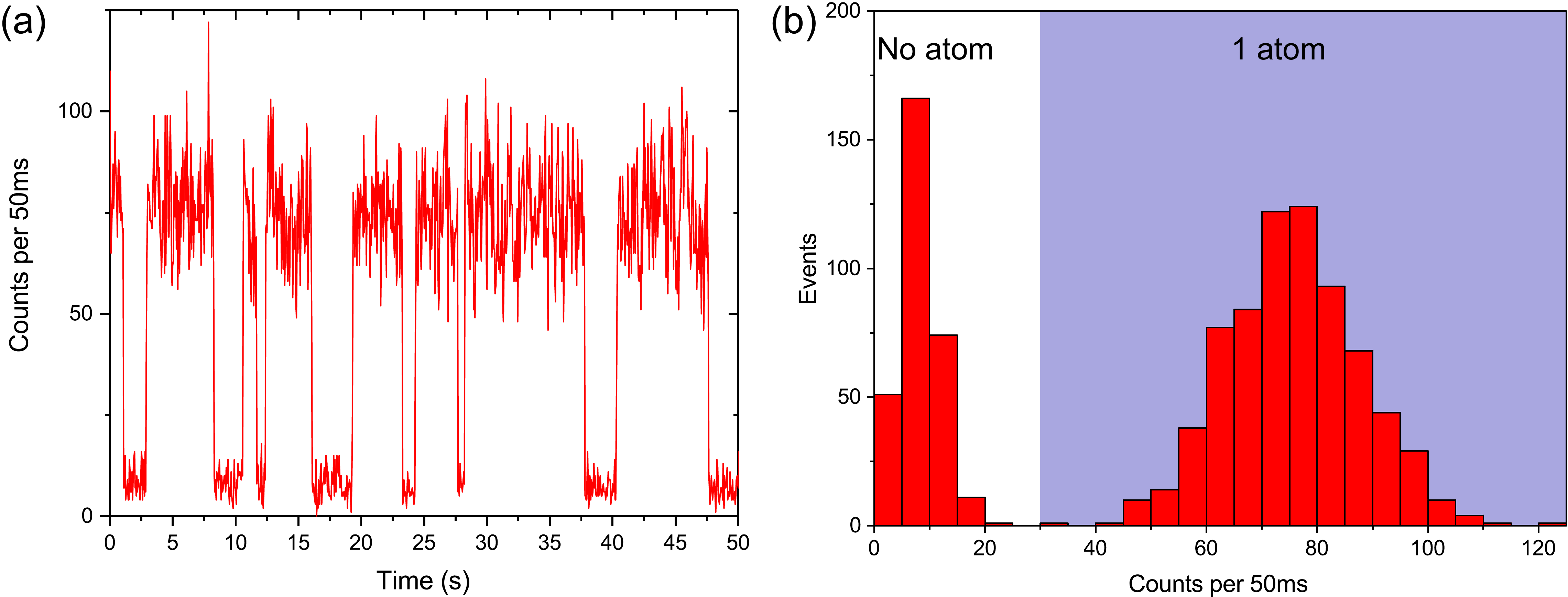}
\caption{\label{fig6} Single atom loading and observation by using the NA=0.29 objective lens. (a) and (b) are the recorded single atom signal and the corresponding histogram.}
\end{figure}

The performance of the NA$=0.4$ objective is also verified by the same setup and procedures as the test of NA$=0.29$ objective. The results are also shown in Fig. \ref{fig5} as (c)-(f). The image of USAF 1951 resolution target is displayed in (c), from which the magnification of imaging system is determined as $-17.68$. We can also see that the line pairs in element 4 of group 8 can be resolved, and this gives the resolution of the objective about 1.4 $\mu$m that is consistent with the theoretical result $0.61 \lambda / \text{NA}= 1.3$ $\mu$m. The FOV of the objective lens is confirmed to be greater than 0.6 mm by the measurements of PSF and Strehl ratio [Fig. \ref{fig5} (d)-(f)].

When the NA=0.29 objective lens in the single atom manipulation setup [Fig. \ref{fig3}(a)] is replaced by this NA=0.4 objective lens, a smaller trap with size about 1.8 $\mu$m can be constructed. The single atom can also be loaded. (a) and (b) in Fig. \ref{fig6} shows the single atom signal and the histogram, respectively. The single atom signal is clearly resolved from the background.

The third plano-concave silica lens (LA4380) can be replaced by the counterpart made from N-BK7 (LA1509). The distance between the first two lenses should be optimized to d=1.6 mm to get the diffraction limited condition. The calculated Strehl ratio versus off-axis displacement is shown in Fig. \ref{fig5}(b) as blue dashed line. The theoretical FOV is then decreased to about 0.5 mm.

\begin{table}[ht]
\caption{\label{tab4} (a) NA$=0.4$ objective optimized for various wavelength with the window thickness d=5 mm. (b) NA$=0.4$ objective optimized for various window thickness D at the wavelength of 852 nm. NA is kept as 0.4 for all the optimizations. $\lambda$, d, and D are the working wavelength, distance between the first two lenses (thickness of surface 2 in Table \ref{tab3}), and window thickness, respectively.}
\centering
\begin{tabular}{l}
(a)  \\
\begin{tabular}{cccc}
\hline \hline
$\lambda$ (nm) & d (mm) & EFL (mm) & FOV (mm)\\
\hline
405 & 10.5 & 25.0 & 0.18 \\
436 & 9.1 & 25.5 & 0.22 \\
486 & 7.2 & 26.3 & 0.29 \\
532 & 5.9 & 26.9 & 0.35 \\
589 & 4.7 & 27.4 & 0.42 \\
671 & 3.3 & 28.0 & 0.50 \\
780 & 2.0 & 28.6 & 0.57 \\
852 & 1.3 & 28.9 & 0.61\\
1060 & 0.9 & 29.2 & 0.69 \\
\hline
\end{tabular}
\\
\\
(b) \\
\begin{tabular}{cccc}
\hline \hline
D (mm) & d (mm) & EFL (mm) & FOV (mm)\\
\hline
7 & 17.3 & 24.2 & 0.6 \\
6 & 7.0 & 27.0 & 0.61 \\
5 & 1.3 & 28.9 & 0.61 \\
\hline
\end{tabular}
\end{tabular}
\end{table}

Like the NA$=0.29$ objective, this NA$=0.4$ objective can also achieve diffraction limit at various working wavelength and window thickness in addition to the case at 852nm and 5-mm vacuum window. The diffraction limited condition is also achieved by only adjusting the distance d between the first two lenses (LB1093 and LB1676). The maximum NA with 0.4 is always maintained for all the optimizations. Table \ref{tab4} (a) and (b) gives the results of optimization.

\section{Conclusion}
In this paper we presented two large NA objective assemblies made from entire commercial on catalog singlet lenses. The key features of two objectives are summarized in Table \ref{tab5}. The performances of the two objectives are simulated by the commercial ray-tracing software and tested experimentally by imaging a USAF 1951 resolution target and 1.18 $\mu$m pinhole with the aid of a achromatic lens. The results show that the two objectives can work with diffraction limit over the full FOV.

\begin{table}[ht]
\caption{\label{tab5} Key features of the objectives. }
\centering
\begin{tabular}{ccccc}
\hline \hline
NA & EFL (mm) & WD (mm) & FOV (mm) & Resolution ($\mu$m)\\
\hline
0.29 & 34.3 & 34.6 & 0.62 & 1.8 \\
0.4 & 28.9 & 18.2 & 0.61 & 1.3 \\
\hline
\end{tabular}
\end{table}

Although that the two objectives are designed for 5 mm silica window at 852 nm, they can also be optimized to work at diffraction limit for various thickness of window at wavelength from visible to near inferred region by only adjusting the distance of first two singlets. The whole structures of the objectives are compatible to the mostly used $\Phi$1'' optics and small glass cells, and thus we believe that they are suitable for imaging and resolving single atoms for various of species and vacuum cells. These objectives not only provide the cold atoms community more cheap and easy-obtained options to resolve and manipulate single atoms, but also could be used for other applications, such as in biophysics to monitoring the living cell or industry to check the small structure over long distance.

\begin{acknowledgments}
This work was supported by the National Key Research and Development Program of China (Grant No. 2017YFA0304502), the National Natural Science Foundation of China (Grant No. 11634008, 11674203, 11574187, and 61227902), and the Fund for Shanxi "1331 Project" Key Subjects Construction.
\end{acknowledgments}

\bibliography{objectivebib}

\end{document}